\documentstyle[epsf,graphicx]{mn}

\def\x2{$\chi^{2}$}

\def\lunits{$\rm erg~s^{-1}$}
\def\funits{$\rm erg~cm^{-2}~s^{-1}$}
\def\cunits{$\rm cm^{-2}$}

\newbox\grsign \setbox\grsign=\hbox{$>$} \newdimen\grdimen \grdimen=\ht\grsign
\newbox\simlessbox \newbox\simgreatbox \newbox\simpropbox
\setbox\simgreatbox=\hbox{\raise.5ex\hbox{$>$}\llap
     {\lower.5ex\hbox{$\sim$}}}\ht1=\grdimen\dp1=0pt
\setbox\simlessbox=\hbox{\raise.5ex\hbox{$<$}\llap
     {\lower.5ex\hbox{$\sim$}}}\ht2=\grdimen\dp2=0pt
\setbox\simpropbox=\hbox{\raise.5ex\hbox{$\propto$}\llap
     {\lower.5ex\hbox{$\sim$}}}\ht2=\grdimen\dp2=0pt

\def\rosat{{\it ROSAT~}}

\def\xmm{{\it XMM-Newton~}}
\def\chandra{{\it Chandra~}}

\begin{document}

\title[XBONGs: more diluted than absorbed?] 
{X-ray Bright Optically Inactive Galaxies in XMM-Newton/SDSS fields:
more diluted than absorbed?}  

\author[Georgantopoulos \& Georgakakis]
{ I. Georgantopoulos \& A. Georgakakis \\
Institute of Astronomy \& Astrophysics, National Observatory of
Athens, I. Metaxa \& B. Pavlou, 15236, Athens, Greece} 

\maketitle
\begin{abstract}
We explore the properties of X-ray Bright Optically Inactive Galaxies
(XBONG) detected in the 0.5-8\,keV spectral band in 20  public \xmm
fields overlapping with the SDSS.  We  constrain our sample to optically 
 extended systems with $\log f_X / f_{opt} > -2$ that have 
spectroscopic identifications available from the SDSS ($r<19.2$\,mag). 
 The resulting sample contains
12 objects with $L_X (0.5-8 \rm \, keV)= 5 \times 10^{41} - 2 \times
10^{44}\, erg  \, s^{-1}$ in the redshift range $0.06 < z < 0.45$. The
X-ray emission in four cases is extended  suggesting the presence of
hot gas associated with a cluster or group of galaxies. The X-ray
spectral fits show that two additional sources are best fit with a
thermal component emission ($\rm kT\sim 1 keV$). Three sources are
most likely  associated with AGN: their X-ray spectrum is  described
by a steep photon index $\Gamma\sim 1.9$ typical of unobscured AGN
while, they are very luminous in X-rays ($L_X (0.5-8 \rm \, keV)
\approx 10^{43} - 10^{44} \, erg \, s^{-1}$. Finally, three more
sources could  be associated with either normal galaxies or unobscured Low
Luminosity AGN  ($\rm L_X < 10^{42}$\,\lunits).  We find no evidence
for significant X-ray absorbing columns in any of our XBONGs.  
The above suggest that XBONGs, selected in the total 0.5-8 keV band,
comprise a mixed bag of objects primarily including normal elliptical
galaxies  and type-1 AGN whose optical nuclear spectrum is probably
diluted by the  strong stellar continuum.  Nevertheless, as our sample
is not statistically complete we cannot exclude the possibility  that
a fraction of optically fainter XBONG  may be associated with heavily
obscured AGN.         
\end{abstract}


\section{Introduction}
 Deep \chandra surveys have resolved the bulk of the X-ray background
 in both the soft and the hard energies  (Mushotzky et al. 2000;
 Brandt et al. 2001; Giaconni et al. 2002; Alexander et al. 2003).  
 A striking result is that these surveys do not find a
 single dominant population of heavily obscured AGNs, predicted by the
 X-ray background population synthesis models. On the contrary a
 heterogeneous population of sources is detected comprising a mix of
 (i) BL AGN (QSOs and Seyfert-1 galaxies), (ii) narrow emission line
  AGN, (iii) optically faint sources ($I>24$\,mag) and (iv) 'passive'
 galaxies with absorption line optical spectra. The latter class of
 sources, frequently dubbed X-ray Bright Optically Inactive Galaxies
 (XBONGs), shows no sign of AGN  activity in their optical spectra
 (e.g. no emission lines), while the X-ray luminosity is large enough 
 ($\rm \ga 10^{42}\, erg \, s^{-1}$) that is hard to reconcile without
 invoking the presence of AGN activity (Fiore et al. 2003;
 Hornschemeier et al. 2001; Barger et al. 2001). Although this
 class of sources has  been detected in previous low resolution  X-ray 
 missions (Elvis et al. 1981; Griffiths et al. 1995; Moran et
 al. 1996; Blair, Georgantopoulos \& Stewart 1997) it is only recently
 that they have  received much attention. This interest has been
 initiated by the  suggestion that these sources host heavily obscured
 AGNs  (e.g. Comastri et al. 2002) and therefore, they may be the
 missing   link between observations and model predictions. Indeed, a
 large  fraction of completely hidden AGNs may be hosted by optically  
 normal galaxies, partially explaining the scarcity of obscured AGN in
 deep X-ray surveys (Comastri et al. 2002). 

 Alternatively, the lack of optical emission lines can be explained if
 the nuclear component is outshined  by the strong stellar continuum
 (e.g. Severgnini et al. 2003). The X-ray spectra provide an
 invaluable tool for discriminating between the above
 possibilities. Unfortunately, the sources detected in the deep  
 \chandra fields  are in general too faint to allow detailed 
 X-ray spectral analysis. 

 Therefore, it is important to find nearby,  bright 
 examples  of XBONGs in wide angle relatively shallower surveys.
 Wide areal coverage is difficult to achieve with the {\it Chandra}
 observatory due its limited field--of--view. On the contrary,
 {\it XMM-Newton} with 4 times larger field-of-view provides an ideal
 platform for such a study. In this paper we exploit the capabilities
 and the large volume of archival data of the XMM-{\it  Newton} to
 serendipitously identify  XBONGs in public fields selected to overlap
 with the Sloan Digital Sky Survey (SDSS; York et al. 2000). This is
 to exploit the superb and uniform 5-band optical photometry and
 spectroscopy available in this area.  Throughout this paper we adopt
 $\rm H_o=65 \,  km \, s^{-1} \, Mpc^{-1}$ and  $\rm \Omega_M=0.3$,
 $\rm \Omega_{\Lambda}=0.7$.

\section{The Data}

\subsection{The SDSS data} 
In this paper we use {\it XMM-Newton} archival observations with a
proprietary period that expired before September 2003 and  that
overlap with the first data release of the SDSS (DR1; Stoughton et
al. 2002). The SDSS is an on-going imaging and spectroscopic survey
that aims to   cover about $\rm 10\,000\,deg^2$ of the sky. Photometry
is performed in 5 bands ($ugriz$;  Fukugita et al. 1996; Stoughton et
al. 2002) to the limiting magnitude $g \approx 23$\,mag, providing a
uniform and  homogeneous multi-color photometric catalogue. The SDSS
spectroscopic observations will obtain spectra for over 1 million
objects, including galaxies brighter than $r=17.7$\,mag, luminous red
galaxies to $z\approx0.45$ and colour selected QSOs (York et al. 2000;
Stoughton et al. 2002).

\subsection{The \xmm data}
The {\it XMM-Newton} archival observations used here have the EPIC
(European Photon Imaging Camera;  Str\"uder et al. 2001; Turner et
al. 2001) cameras as the prime instrument operated in full frame
mode. For fields observed more than once with the {\it XMM-Newton} we
use the deeper of the multiple observations. The total of 20 XMM-{\it
Newton} fields used here are listed in Table \ref{log}.  

The X-ray data have been analysed using the Science Analysis Software
({\sc sas}  5.4). The event files produced by the {\it XMM-Newton}
Science Center data reduction pipeline were screened for high particle
background periods by rejecting time intervals with 0.5-10\,keV count
rates higher than 30 and 15\,cts/s for the PN and the two MOS cameras
respectively.  The PN and MOS good time intervals for these pointings
are shown in Table  \ref{log}. The differences between the PN and MOS
exposure times are due to varying start and end times of individual
observations. Only events  corresponding to patterns  0--4 for the PN
and 0--12 for  MOS have been kept. 

In order to increase the signal--to--noise ratio and to reach fainter
fluxes the PN and the MOS event files have been combined into a single
event list using the {\sc merge} task of SAS. Images have been
extracted in the spectral bands 0.5-8 (total), 0.5-2 (soft) and
2-8\,keV (hard) for both the merged and the individual PN and MOS
event files. We use the  more sensitive (higher S/N ratio) merged
image for source  extraction and flux estimation, while the individual
PN and MOS images are used to calculate hardness ratios. This is
because the interpretation of hardness ratios is simplified if the
extracted count rates  are from one detector only.  Exposure maps
accounting for vignetting, CCD gaps and bad pixels  have been
constructed for each spectral band. In the present study the source
detection is performed on the 0.5-8\,keV image using the {\sc
ewavelet} task of SAS with a detection threshold of  $5\sigma$. In
this paper we only consider sources with offaxis angles
$<13.5$\,arcmin and therefore the total surveyed area is about $\rm
3.14\,deg^2$. A total of 1286 X-ray sources have been detected to
the limit  $f_X(\rm 0.5-8\,keV) \approx 2\times10^{-15}\, erg \,
s^{-1} \, cm^{-2}$. About 10 per cent of the total survey area is
covered at the flux limit $f_X(\rm 0.5-8\,keV) \approx
3\times10^{-15}\, erg \, s^{-1} \, cm^{-2}$. This fraction increases
to about 50 per cent at $\rm \approx 7 \times 10^{-15} \, erg \,
s^{-1} \, cm^{-2}$.

Count rates in the merged (PN+MOS) images as well as the
individual PN and MOS images are estimated within an 18\,arcsec
aperture. For the  background estimation we use the background maps
generated as a by-product of the {\sc ewavelet} task of  SAS. A small
fraction of sources lie close to masked regions (CCD gaps or hot
pixels) on either the MOS or the PN detectors. This may introduce
errors in the estimated source  counts. To avoid this bias, the source
count rates (and hence the hardness ratios and the flux) are estimated
using the detector (MOS or PN) with no masked pixels in the vicinity
of the source. 

We convert counts to flux assuming a power-law spectrum with
$\Gamma=1.7$ and the appropriate Galactic absorption for each field
listed in Table \ref{log} (Dickey \& Lockman 1990).  The mean count-rate 
 to flux conversion (or Enery Conversion Factor, ECF) for
the mosaic of all three detectors is estimated by weighting the ECFs
of individual detectors by the respective exposure time.  For the
encircled energy correction, accounting for the energy fraction
outside the aperture within which source counts are accumulated, we
adopt the calibration given by the {\it XMM-Newton} Calibration
Documentation  
\footnote{http://xmm.vilspa.esa.es/external/xmm\_sw\_cal/calib \\
/documentation.shtml\#XRT}.
  
\begin{table*}
\footnotesize
\begin{center}
\begin{tabular}{cc cc cc l}
\hline
 RA  & Dec  & FILTER & $\rm N_H$  & PN exp. time & MOS1 exp. time &
 Field name \\

(J2000) & (J2000) & & ($\rm 10^{20}\,cm^{-2}$)  & (sec) &  (sec) &  \\
\hline
23 54 09 & --10 24 00 & MEDIUM & 2.91 & 13\,600 & 19\,100 & ABELL\,2670 \\
23 37 40 & +00 16 33 & THIN   & 3.82 & 8\,200  & 13\,300 & 
RXCJ\,2337.6+0016 \\
17 01 23 &  +64 14 08 & MEDIUM & 2.65 & 2\,300  & 3\,900  & RXJ\,1701.3 \\
15 43 59 &  +53 59 04 & THIN   & 1.27 & 14\,200 & 19\,200 & SBS\,1542+541 \\
13 49 15 &  +60 11 26 & THIN   & 1.80 & 14\,100 & 18\,100 & NGC\,5322 \\
13 04 12 &  +67 30 25 & THIN   & 1.80 & 14\,600 & 17\,100 & ABELL\,1674 \\
12 45 09 & +00 27 38 & MEDIUM & 1.73 & 46\,300 & 55\,500 & NGC\,4666 \\
12 31 32 &  +64 14 21 & THIN   & 1.98 & 26\,100 & 30\,100 & MS\,1229.2+6430 
\\
09 35 51 &  +61 21 11 & THIN   & 2.70 & 20\,400 & 33\,900 & UGC\,5051 \\
09 34 02 &  +55 14 20 & THIN   & 1.98 & 23\,500 & 28\,500 & IZW\,18 \\
09 17 53 &  +51 43 38 & MEDIUM & 1.44 & 15\,900 & 13\,600 & ABELL\,773 \\
08 31 41 &  +52 45 18 & MEDIUM & 3.83 & 66\,800 & 73\,300 & APM\,08279+5255 
\\
03 57 22 &  +01 10 56 & THIN   &13.20 & 19\,100 & 21\,400 & HAWAII\,167 \\
03 38 29 &  +00 21 56 & THIN   & 8.15 & 8\,900  & 6\,700  & 
SDSS\,033829.31+00215 \\
03 02 39 &  +00 07 40 & THIN   & 7.16 & 38\,100 & 46\,900 & CFRS\,3H \\
02 56 33 &  +00 06 12 & THIN   & 6.50 &  --     & 11\,600 & RX\,J0256.5+0006 
\\
02 41 05 & --08 15 21 & MEDIUM & 3.07 & 12\,300 & 15\,600 & NGC\,1052 \\
01 59 50 & +00 23 41 & MEDIUM & 2.65 & 3\,800  &  --     & MRK\,1014 \\
01 52 42 &  +01 00 43 & MEDIUM & 2.80 & 5\,800  & 17\,200 & ABELL\,267 \\
00 43 20 & --00 51 15 & MEDIUM & 2.33 & 15\,700 &  --     & UM\,269 \\
\hline
\end{tabular}
\end{center}
\caption{The archival {\it XMM-Newton} pointings used in this study.
}\label{log}
 \normalsize
\end{table*}

\begin{figure*}
\rotatebox{0}{\epsfxsize=18cm \epsffile{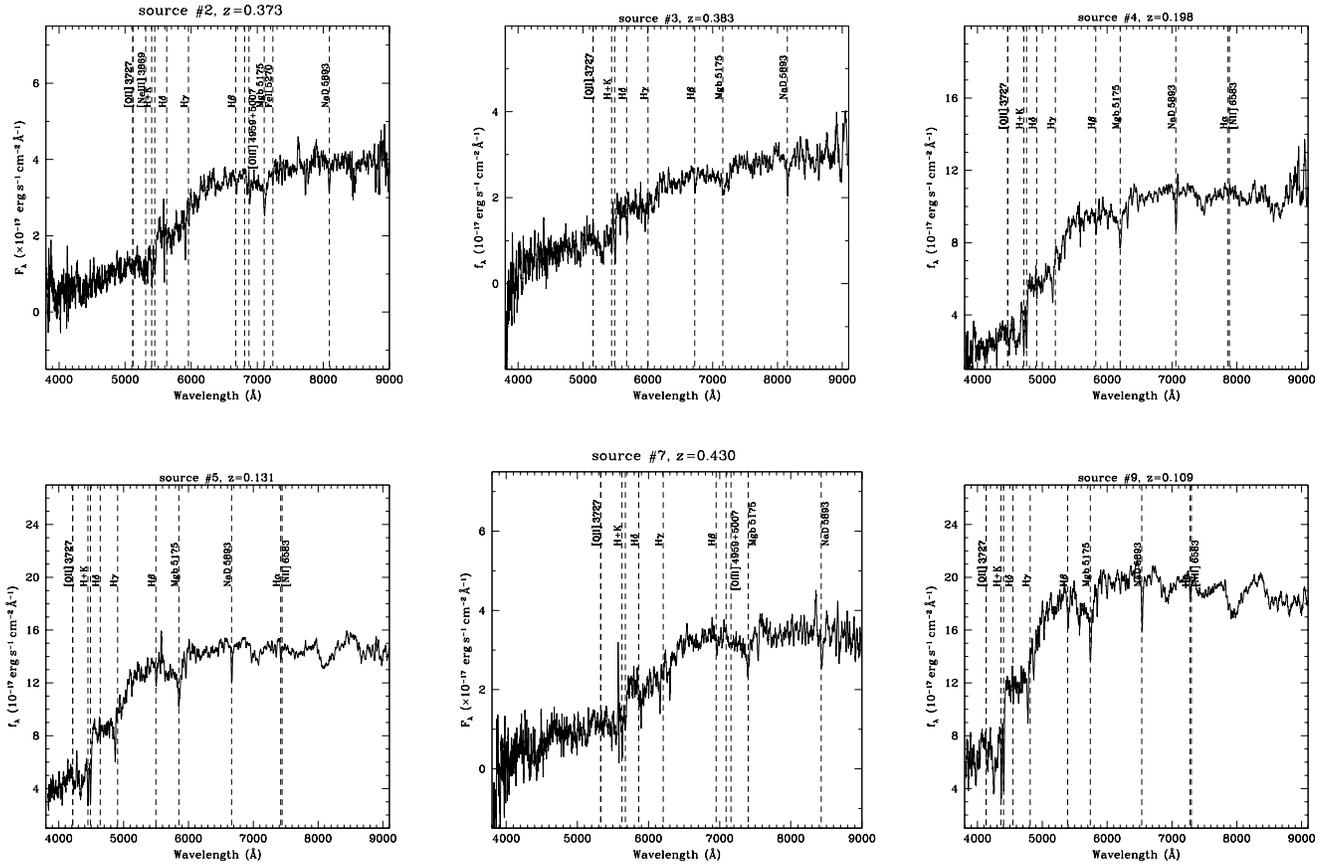}}   
\caption{
 The optical spectra of the six sources that are likely to 
 be associated with AGN. 
}\label{spec_opt} 
\end{figure*}

\section{The sample} 

 The SDSS optical photometric catalogue is used to identify optical
 counterparts to the X-ray sources by estimating the probability, $P$,  
 that a given candidate is a chance coincidence (Downes et al. 1986).
 The probability depends on both the separation of the optical
 counterpart from the X-ray centroid and the surface density of the
 optical sources at the given magnitude (see Georgakakis et al. 2004
 for details). For our identifications we adopt a  probability
 threshold $P<0.015$ and a maximum search radius of 7\,arcsec. 

\begin{figure}
\rotatebox{0}{\epsfxsize=3in \epsffile{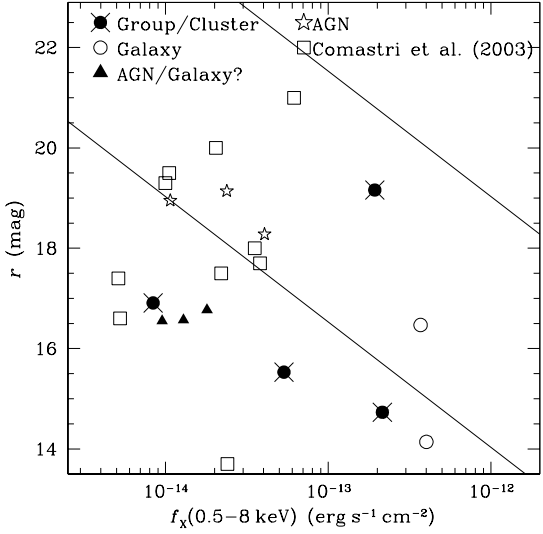}}   
\caption{The X-ray to optical flux diagram for the 12 XBONGs in our
 sample compared to the 10 XBONGs from Comastri et
 al. (2003). The solid lines indicate constant 
 X-ray--to--optical flux ratios of $\log f_X/f_{opt}=\pm1$  and
 delineate the region of the parameter space occupied by powerful
 AGNs.}    
 \label{fxfo}
\end{figure}

 We are further selecting sources which  (i) are associated with
 galaxies i.e. they are optically extended sources according to the  
 SDSS star-galaxy classification and (ii) have optical SDSS spectra 
 available. A total of 33 sources fulfill the above criteria of
 which 14 have `early-type' spectra showing absorption lines only
 and lacking emission lines. Finally, we exclude two sources which
 have low X-ray to optical flux ratio, $\log f_X/f_{opt}<-2$ and
 therefore are most probably associated with 'normal' galaxies 
 i.e. their X-ray emission comes from binaries and diffuse hot gas (see
 Hornschemeier et al. 2003). Both these sources are detected at the vicinity 
 of the cluster Abell2670, having a redshift of z=0.07, and 
 and are most likely   associated with the above cluster. 
 Their X-ray coordinates (J2000) are $\alpha=23^h53^m~40.5^s,~\delta=-10^d~24^m~20^s$
 and $\alpha=23^h54^m~5.7^s,~\delta=-10^d~18^m~31^s$. 
 Their luminosities in the 0.5-8 keV band are $3\times10^{41}$ and $10^{41}$ 
 \lunits respectively. The optical and X-ray properties of the remaining
 12 sources are presented in Table \ref{opt}. 
 The optical spectra of the six sources which are likely to 
 be associated with AGN (see the discussion section) are shown in
 Fig. \ref{spec_opt}. Clearly our selection criteria do not provide a 
 complete sample of XBONGs that can be used for statistical studies
 (e.g. surface density of XBONGs). Nevertheless, our galaxies span a
 range of X-ray--to--optical flux ratios (see Fig. \ref{fxfo}) similar
 to those probed by other XBONG samples compiled from \xmm and
 \chandra surveys (Comastri et al. 2003), suggesting that they are
 representative of the XBONG population.      

 Four of our sources (\#1, 6, 8, 10) are extended on the \xmm images.
 The angular extensions of these sources correspond to 80-500
 kpc implying that the X-ray emission comes from hot   
 intracluster medium. Three more sources are galaxy cluster members. 
 Source \#2 at a redshift of  $z=0.373$ is
 associated with a cluster also detected by \rosat  
 while sources 11 and 12 at a redshift
 $z\sim0.07$ are associated with the cluster Abell\,2670; all three
 sources are embedded in the diffuse X-ray emission of the clusters.
 We note that the fluxes and luminosities for the above sources
 may present errors larger than those implied by 
 photon statistics alone because of uncertainties in 
 the background subtraction.

\begin{table*}
\scriptsize
\begin{center}
\begin{tabular}{ccccccccccccc}
\hline
\# &RA     &Dec  & $z$ & Offset & $r$ & 
 $f_X^3$& HR & $\log
 f_X/f_{opt} $& $L_X^4$ & $M_r$ & $\rm Class^5$ & Field \\ 

 &(J2000)&(J2000)& &(arcsec)& &     
 &  &       &    &  
 &       \\
\hline 
1 & 01 53 15.2 & $+01$ 02 20 & 0.060 & 0.3 & 14.73 &
$21.5$ & $-0.93\pm0.06^2$ & -1.39 & 2.2 & -22.58 & Group & ABELL267 \\

2 & 02 56 30.8 & $+00$ 06 02 & 0.373 & 1.3 & 18.28 &
$4.1$ & $-0.41\pm0.07^2$ & -0.21 & 21.5 &-23.40 & AGN & RXJ0256.5  \\

3 & 03 01 53.9 & $+00$ 15 37 & 0.383 & 3.5 & 18.95 &
$1.1$ & $-0.15\pm0.27^1$ & -1.00 & 6.0 & -22.80 & AGN & CFRS3H \\

4& 03 38 10.0 & $+00$ 16 10 & 0.198 & 1.6 & 16.77 & 
$1.8$ & $<-0.34^1$  & -1.65 & 2.3 & -23.35 & AGN/Gal? & SDSS033829 \\

5 & 09 36 19.4 & $+61$ 27 21 & 0.131 & 2.3 & 16.55 &
$0.95$ & $-0.76\pm1.0^2$ & -2.00 & 0.52 & -22.62 & AGN/Gal? & UGC5051 \\

6 & 12 44 54.5 & $-00$ 26 39 & 0.231 & 0.8 & 16.91 &
$0.84$ & $-0.67\pm0.20^2$ & -1.93 & 1.5 & -23.60 & Group & NGC4666 \\

7 & 13 03 28.8 & $+67$ 26 41 & 0.430 & 2.5 & 19.14 &
$2.4$ & $-0.83\pm0.19^1$ & -0.59 & 17.3 & -22.89 & AGN & ABELL1674 \\

8 & 13 02 40.3 & $+67$ 28 42 & 0.106 & 1.1 & 15.53 &
$5.3$ & $-0.59\pm0.13^1$ & -1.67 & 1.9 & -23.16 & Group & ABELL1674 \\

9 & 13 02 51.0 & $+67$ 25 17 & 0.109 & 5.0 & 16.57 &
$1.3$ & $-0.56\pm1.0^1$  & -1.87 & 0.46 & -22.13 & AGN/Gal? & ABELL1674 \\

10 & 17 01 23.8 & $+64$ 14 13 & 0.452 & 1.8 & 19.16 &
$19.3$ & $-0.48\pm0.11^2$ & 0.34 & 159.7 & -23.02 & Group & RXJ1701.3 \\

11 & 23 54 07.2 & $-10$ 25 15 & 0.071 & 3.0 & 16.47 &
$36.9$ & $-0.69\pm0.04^2$ & -0.46 & 5.2 & -21.23 & Gal & ABELL2670 \\

12 & 23 54 13.9 & $-10$ 25 09 & 0.078 & 2.8 & 14.14 &
$40.1$ & $-0.52\pm0.04^2$  & -1.35 & 6.9 & -23.77 & Gal & ABELL2670 \\
\hline 
\multicolumn{11}{l}{$^1$ PN, $^2$ MOS, $^3$ in units $10^{-14}$ \funits (0.5-8 keV), 
$^4$ in units $10^{42}$ \lunits (0.5-8 keV)$^5$ Classification based on X-ray properties 
}
\end{tabular}
\end{center}
\caption{Optical and X-ray properties of the 12 XBONGs identified in
this study}
\normalsize
\label{opt}
\end{table*}

\section{The X-ray spectra} 

\subsection{Hardness Ratios} 
\begin{figure}
\rotatebox{0}{\epsfxsize=6.5cm \epsffile{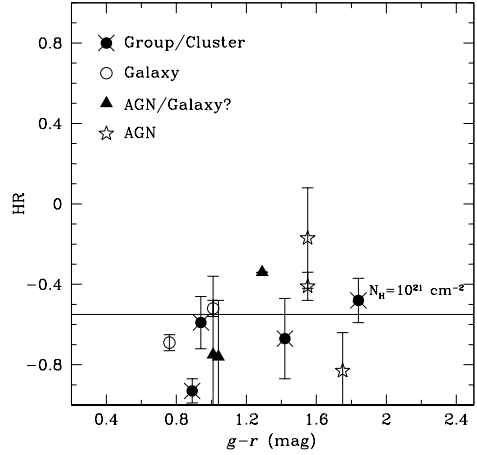}}   
\caption{The hardness ratios  as a function of the colour $g-r$. 
 The solid line denotes a power-law spectrum with
 $\Gamma=1.9$ and absorption $\rm N_H=10^{21}\, cm^{-2}$ for the MOS  
 detector. The error bars correspond to the 1$\sigma$ confidence
 level. The point with no error bar corresponds to a 3$\sigma$ upper limit
 (source 4).
 }   
 \label{hr}
\end{figure}

In Fig. \ref{hr} we plot the hardness ratio as a function of the $g-r$
colour. The hardness ratio, $\rm HR$, is  defined as
\begin{equation}\label{eq1}
\rm HR = \frac{\rm RATE(2080)-RATE(0520)}{\rm RATE(2080)+RATE(0520)},
\end{equation}
\noindent 
where $\rm RATE(0520)$ and $\rm RATE(2080)$ are the count rates in
the 0.5-2 and 2-8\,keV spectral bands respectively. 
For one source (4), we have less than 5 counts in the hard band 
and thus we can estimate only an 3$\sigma$ upper limit 
 on the hardness ratio. 
The hardness ratios are
estimated using the PN data except for sources that lie close
to PN CCD gaps or hot pixels where we use MOS data (see section
2.2). These sources are marked in Table 2.   

 For comparison in Fig. \ref{hr} we also plot the
 line which corresponds to a power-law spectrum with photon index
 $\Gamma=1.9$ and an absorbing column of $\rm N_H = 10^{21} \,
 cm^{-2}$ for the MOS detector. Note however, that as the MOS has a
 lower effective area at low energies, compared to the PN, the same
 hardness ratio corresponds to slightly different $\rm N_H$
 values. The vast majority of sources present soft spectra,  
 with values of the hardness ratio  clustering around $-0.5$.
  All sources  present red
 colours ($g-r>0.5$) typical of early-type systems.     

\begin{figure}
\rotatebox{270}{\epsfxsize=4in \epsffile{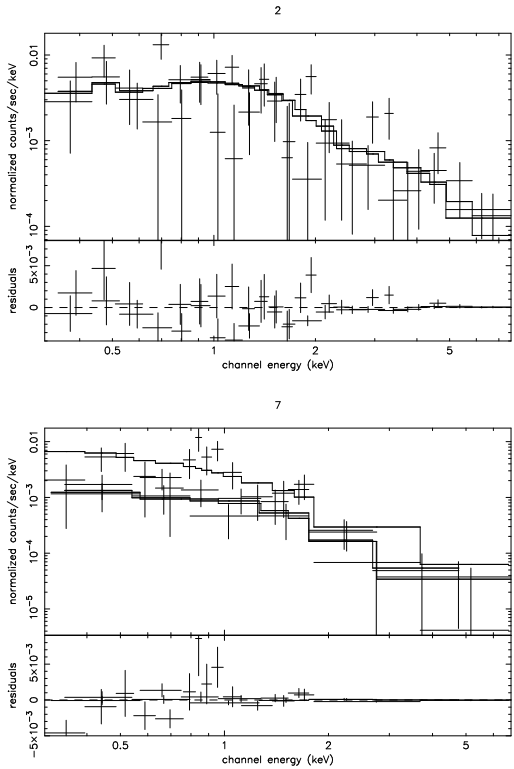}}   
\caption{
 X-ray spectra of two candidate AGN  (\#2
 and 7) The power-law fit to the data is shown as well  
 together with the corresponding residuals. 
 }
 \label{xspec}
\end{figure}

\subsection{Spectral Fits} 
Next, we attempt to constrain the spectral properties of the subsample
of the 9 sources in Table \ref{opt} (sources \#1, 2, 3, 6, 7, 8, 10,
11, 12) that have sufficient counts to perform X-ray spectral
analysis. For the remaining sources (\#4, 5, 9) poor photon statistics
do not allow spectral fittings to be performed. The source spectra are
extracted using a radius of 18\,arcsec. In most cases, the background
spectrum is estimated from nearby image regions free from sources. In
the case of the sources \#2, 11 and 12 which are embedded in strong
cluster emission, the background was taken from adjacent regions in
the cluster. Response matrices and auxiliary files are generated using
the {\sc sas} tasks {\sc rmfgen} and {\sc arfgen} respectively. We
use the {\sc xspec} v11.2 software to perform  the spectral fits. The
quoted errors correspond to the 90 per cent confidence level. The
spectral fits are performed in the 0.3-8\,keV band where the
instruments calibration are well known. In the case of the sources
\#1, 6, 10, 11 and 12 we have a sufficient number of counts (more than
15 counts per bin) to perform $\chi^2$ statistics. In the other cases
 (\# 2, 3 7,6) where much fewer counts are observed, we use the C-statistic instead
(Cash 1979) which does not require for the binning of the data. The
disadvantage is that the C-statistic does not allow for the derivation
of the goodness-of-fit probability  unlike the $\chi^2$ statistic. 
      
We fit two models to the spectra: a Raymond-Smith model which provides
a good representation of the hot gas emission  in early-type normal
galaxies, and a power-law model which is characteristic of AGN
spectra. The spectral fit results are given in Table 3. Examples of
the X-ray spectra of the two X-ray brighter sources in the sample (not
associated with extended X-ray emission) are shown in Figure
\ref{xspec}. The values with no errors denote that the parameter is
tied during the fitting. The 
column density is fixed in most cases to the Galactic  value (Dickey
\& Lockman 1990). In the case of the  Raymond-Smith model, the
abundance is fixed to $Z=0.3$. Comparison of the resulting
$\chi^2$ values of the Raymond-Smith and the power-law models
(Mushotzky 1982) in the case of sources \#1, 11 and 12 demonstrates
that the Raymond-Smith  spectrum provides a significantly better fit
compared to the power-law spectrum. The best fit temperatures are
$\sim1$\,keV typical for early-type galaxies or groups (Matsumoto et
al. 1997). This strongly suggests that the X-ray emission  in these
objects is associated with hot gas emission.
 Sources \#11 and 12 are not extended and thus are likely 
 associated with gas heated by the gravitational potential of a  galaxy.  
Sources \#1 as well as  6, 8, 10 are extended and thus most likely associated  with
groups (\#1, 6, 8) or clusters of galaxies (\#10). The best fit
temperature $\rm kT\sim 1-3\,keV$ and luminosities $\rm L_X (\rm 0.5 -
8 \, keV) \sim 10^{42} - 10^{44} \, erg \, s^{-1}$ are consistent with
the above interpretation. The spectral fits for sources \#2, 3  and 7
are good  for both the Raymond-Smith and the power-law models.   
Therefore, these could be in principle associated with either normal
galaxies or unobscured AGN. However, the best fit temperatures are
higher (2-5\,keV) than those of normal galaxies favoring, together
with high X-ray luminosities, the AGN case.

\begin{table*} 
\caption{Spectral fits to the 9 brighter sources}
\begin{tabular}{ccccccc}
\hline 
Object   & \multicolumn{3}{c}{Raymond-Smith} & \multicolumn{3}{c}{Power-Law} \\
\hline 
         &  $\rm N_H^1$ & $kT$ & $\chi^2$   & $\rm N_H^1$ & $\Gamma$ & $\chi^2$ \\
\hline 
1 & $9^{+13}_{-5}$ & $0.91^{+0.05}_{-0.03}$& 97.2/92& 3 & $2.2^{+0.1}_{-0.1}$ & 400.9/94\\
2 & 7 & $5.3^{+2.3}_{-1.7}$ & -  & 7 & $2.01^{+0.20}_{-0.30}$ & - \\
3 & 7 & $6.8^{+20.}_{-4.4}$ & -  & 7 & $1.70^{+0.55}_{-0.50}$ & - \\
6 & 1.7 & $1.62^{+0.42}_{-0.32}$&  113.2/107 & 1.7 & $2.29^{+0.18}_{-0.18}$ &120.9/107 \\  
7& 2 & $2.18^{+0.95}_{-0.62}$ & - & 2 & $2.20^{+0.30}_{-0.27}$ & - \\
8& 2 & $2.20^{+1.00}_{-0.60}$ & - & 2 & $2.10^{+0.17}_{-0.22}$ & - \\
10 & 3 & $4.83^{+1.99}_{-2.03}$ & 43.6/63 & 3 & $1.77^{+0.14}_{-0.14}$ & 42.8/63 \\
11& 3 & $1.10^{+0.04}_{-0.04}$ & 553/594 &  3 & $2.22^{+0.09}_{-0.21}$ & 716/594 \\ 
12& 3 & $1.34^{+0.71}_{-0.24}$ & 451/662 & 3 & $2.06^{+0.21}_{-0.21}$ & 468/662 \\
\hline 
\multicolumn{7}{l}{$^1$ Intrinsic rest-frame column density in units
of $10^{20}$ \cunits} \\   
\end{tabular}
\label{}
\end{table*}

\begin{figure}
\rotatebox{0}{\epsfxsize=6.5cm \epsffile{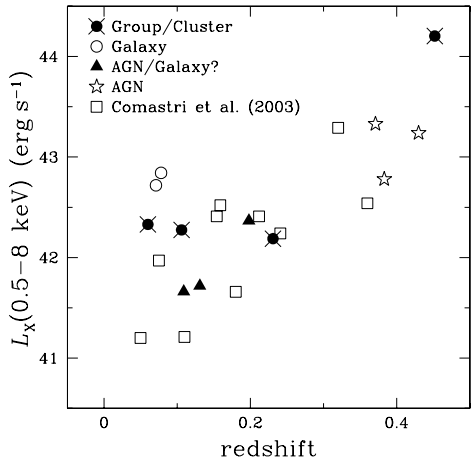}}   
\caption{The luminosity - redshift diagram  for the XBONGs in our sample
compared to those from Comastri et al. (2003).}\label{lz}
\end{figure}

\begin{figure}
\rotatebox{0}{\epsfxsize=3in \epsffile{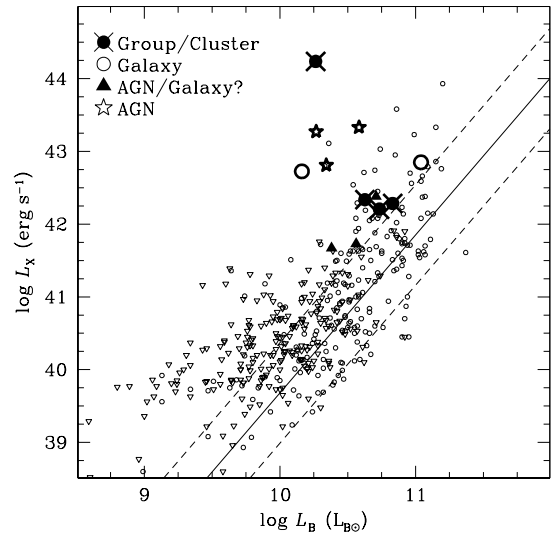}}   
\caption{The $L_X-L_B$ relation for our galaxies  (large symbols) in
comparison with the early-type galaxy sample presented by O'Sullivan,
Forbes \& Ponman (2001). The small open circles and triangles represent 
detections and upper limits in the O'Sullivan et
al. (2001) sample. The continuous line is the best fit to the
O'Sullivan et al.  (2001) data excluding the AGNs, Bright Cluster
Galaxies and dwarfs. The dashed lines are the $1\sigma$ envelopes
around the best fit.  
}\label{lxlb}
\end{figure}

\section{DISCUSSION} 
In this paper we explore the nature of the class of X-ray Bright
Optically Inactive galaxies identified in public {\it XMM-Newton} 
fields that overlap with the SDSS. Our sample is selected in the
0.5-8\,keV spectral band and comprises a total of 12 systems with
$\log f_X / f_{opt} > -2$.  These galaxies represent only a small
fraction (1 per cent) of the full 0.5-8\,keV selected catalogue
totaling 1286 detections. One should bear in mind however, that our
XBONGs do not constitute a complete sample as we selected only the
optically brighter sources with available SDSS spectra. 

A total of 8 sources are indeed, brighter than $r<17.7$\,mag, the
limit of the SDSS galaxy spectroscopic survey, while the remaining 4
have $17.7<  r < 19.2$\,mag and are observed as part of the SDSS
Luminous Red  Galaxy sample. This may suggest that our sample may
preferentially probe the XBONGs with a low $f_x/f_{opt}$. However,
comparison of the  $\log f_X/f_{opt}$ of our objects with the
{\it complete} XBONG samples compiled from the XMM1dF (Fiore et
al. 2003) and the \chandra SSA13  (Barger et al. 2001) surveys,
presented by Comastri et al. (2003), does not support this
claim. Indeed, as shown in Figure \ref{fxfo} our sources  span the same 
range of $\log f_X/f_{opt}$ with the above XBONG samples.

Figure \ref{lz} plots the luminosity--redshift diagram for the XBONGs
in the above surveys in comparison with our sample. The region probed
by our sources ($10^{41} < L_X < 10^{44}\rm \, erg \, s^{-1}$,
$z<0.45$) is comparable to that probed by the objects presented in
Comastri et al. (2003). The evidence above suggests that our 12
objects, although not a complete sample,  provide a fair
representation of the overall XBONG population.    
   
Next we explore the origin of the X-ray emission in our sources and
find that different mechanisms are in operation in different groups
of objects. At least four objects (\#1, 6, 8, 10) show extended X-ray  
emission, ($>80$\,kpc) with luminosities $\sim10^{42} - 10^{44}\rm \,
erg \, s^{-1}$ suggesting diffuse hot gas from a cluster or group of
galaxies. For the remaining sources the  X-ray spectral properties
provide constraints on their nature. In the case of sources \#11,
12 (associated with the cluster Abell\,2670) we find that a
Raymond-Smith model provides a much better fit to the data compared to
a power-law spectral energy distribution.  The derived temperatures
are $\sim1$\,keV, as expected for the hot gas emission encountered in
the weak gravitational potentials of early-type  galaxies. The X-ray 
luminosities are high $5-7\times 10^{42}\rm \, erg \, s^{-1}$ for
thermal emission. However, such high levels of X-ray emission can be
encountered in massive ellipticals especially those residing in
clusters  (eg. Paolillo et al. 2003; O'Sullivan, Ponman \& Collins
2003). This is demonstrated in Figure \ref{lxlb} plotting the $L_X -
L_B$ for the XBONGs in our sample in comparison with the best fit
relation for early type galaxies (excluding AGNs, dwarfs and Bright
Clusters Galaxies; O'Sullivan et al. 2001). Source \#11 ($L_B= 1.4
\times 10^{10} L_{B\odot}$) lies well  above the galaxy $L_X - L_B$
$1\sigma$ locus. This is not surprising as this source is associated
with the central galaxy of the cluster and these often present
enhanced X-ray emission (O'Sullivan et al. 2003). Source \#12 with
$L_B\approx10^{11} L_{B\odot}$ lies marginally above the $1\sigma$
envelope in  Figure \ref{lxlb}.
Three additional sources (\#4, 5, 9) may be associated with normal galaxies
 i.e. the X-ray emission may come from X-ray binaries and hot gas 
 rather than a supermassive black hole. In the case
of these sources the poor photon statistics do not allow us to
derive spectral constraints from the X-ray spectral fitting. However,
the hardness ratios of these objects suggest soft spectra.  The X-ray
luminosities of these objects are relatively low $<2\times
10^{42}$\,\lunits, consistent  with those of normal galaxies or low
luminosity  AGN (eg. LINERS).

Sources \#2, 3, 7  present soft X-ray spectra consistent with either a
power-law or a Raymond-Smith spectrum.  The power-law photon index has
a value of  $\Gamma\approx 2$ typical of an unobscured AGN
spectrum. The  best-fit temperatures in the  case of a Raymond-Smith
model are relatively high (2-5\,keV) more typical of galaxy clusters
than normal galaxies. The lack of extended X-ray emission  lends no
support to the galaxy cluster possibility. Furthermore, the high X-ray
luminosities, $\sim 10^{43} - 10^{44}\rm \, erg \, s^{-1}$ render the
AGN scenario very likely. This is also demonstrated in Figure \ref{lxlb}
where the above three XBONGs deviate from the  early type galaxy $L_X
- L_B$ relation suggesting either AGN or cluster emission. In this
case the absence  of optical  emission lines is  puzzling. 
One possibility could be that the ionizing optical continuum 
 is weak. We have estimated the 3$\sigma$ upper limits for the $H\alpha$ 
 flux using the continuum values in the SDSS spectra for all the 
 six candidate AGNs. In the case where the $H\alpha$ is outside the
 spectral range, we estimate instead the $H\beta$ upper limit and 
 we convert to $H\alpha$ using the standard Balmer decrement value of 3.1.
 We find that in all cases the derived $H\alpha$ to $L_X$ upper limits 
 are in fair agreement with the $H\alpha$ to $L_X$ ratios
 estimated for Broad Line AGN (Ward et al. 1988), lending 
 no support to the weak ionizing continuum case.
Alternatively, dilution of the nuclear light by the  powerful host galaxy  could
explain the lack of emission lines. Severgnini et al. (2003) studied a few XBONGs
in the {\it XMM-Newton} Bright Serendipitous Source sample. They find
a mixture  of obscured ($\rm N_H > 10^{22} \, cm^{-2}$) and unobscured
($\rm N_H < 10^{22} \, cm^{-2}$) AGN with luminosities in the range
$L_X (2-10 \rm \, keV) \sim 10^{42} - 10^{43}$\,\lunits.  They
attribute the lack of emission lines to the faintness of the nucleus
with respect to the galaxy. In particular, they derive that 
 the optical nuclear flux of an AGN with an X-ray luminosity of $5\times10^{42}$ \lunits 
 can be easily outshined by a host galaxy brighter than $M_R=-22$. 
 This scales to $M_R+logL_X<20.7$ for the object to be diluted. 
 All of our AGN candidates satisfy the above criterion, apart from 
 \#2. This source is associated with a large elliptical galaxy
 ($M_r=-23.4$) in a galaxy cluster.
 We cannot then completely exclude the possibility that this source 
 is associated with hot gas emission from either an exceedingly luminous
 galaxy or a cluster subclump.      


Alternative  scenarios for the observed X-ray emission in sources \#2,
3, 4, 5, 7, 9 include (i) BL-Lac type activity (e.g. Blair et
al. 1997) and (ii)  Advection Dominated Accretion Flows (ADAF; Narayan 
\& Yi 1995; Di Mateo et al. 2000). In the case of BL-Lacs one would
expect strong radio emission. However, none of our sources is detected at 
radio wavelengths to the limits of the FIRST (1\,mJy; Faint Images of
the Radio Sky at Twenty centimeters; Becker et al. 1995) or the NVSS
(2.5\,mJy; NRAO-VLA Sky Survey; Condon et al. 1998)  surveys. 
 Nevertheless, one of our sources (\#7; not detected in
the NVSS) has radio--to--optical ($\alpha_{RO}$) and
X-ray--to--optical   ($\alpha_{XO}$) two point correlation  indices
(Stocke et al. 1991) $\alpha_{RO}<0.57$ and $\alpha_{XO}=1.01$ (for
details on the estimation see Georgakakis et al. 2004). On the basis
of the  Stocke et al. (1991) classification scheme the  $\alpha_{RO}$
upper limit and $\alpha_{XO}$ value above  may suggest BL-Lac activity
with radio emission below the NVSS flux density limit. The
$\alpha_{XO}$ and $\alpha_{RO}$ values for the remaining sources are
inconsistent with BL-Lac type activity. Fossati et al. (1998) showed
that the peak frequency of the BL-Lac SED is a function of the radio
luminosity of these objects: less luminous  systems peak at UV and
soft X-ray wavelengths (High energy peak BL-Lacs), while more luminous
radio sources have a peak at the infrared regime (Low energy peak
BL-Lacs). Source \#7 at $z=0.430$ has an upper limit radio luminosity
at 5\,GHz of  $\nu L_\nu<4\times10^{40}\,\rm erg\,s^{-1}$ suggesting
that this (if a BL-Lac) is a  High energy peak BL-Lac (see Figure 7a
of Fossati et al. 1998). For the remaining sources  (\#2, 3, 4, 5, 9)
The estimated radio luminosity upper limit is  $\nu L_\nu \la 5
\times 10^{39}\,\rm erg\,s^{-1}$, lower than the less luminous systems
in the  Fossati et al. (1998) sample. 

 ADAFs coupled with outflows and winds (Blandford \& Begelman 1999) have
 been proposed to explain the hard X-ray emission of nearby elliptical
 galaxies (Allen, Di Matteo \& Fabian 2000).
 We roughly estimate the ratio of the bolometric to the Eddington 
 luminosity $\rm L_{BOL}/L_{EDD}$ for our six candidate AGNs. 
 We assume that the bolometric luminosity is an order of 
 magnitude larger than the X-ray luminosity (e.g. Nicastro et al. 2003). 
 The black hole mass and thus the Eddington luminosity 
 is estimated from the $\rm M_\odot-L_B^{bulge}$ luminosity relation of 
 Gebhardt et al. (2000); the implicit assumption here is that 
 the all six AGN are associated with early-type systems and thus 
 $\rm L_B^{bulge}\approx L_B$. We find that 
 $\rm L_{BOL}/L_{EDD}\sim 10^{-4}-10^{-2}$ and hence
 in principle some of our objects could be associated with ADAFs.       
 However, the ADAF models predict relatively hard X-ray spectra 
 ($\Gamma\approx1.4$) that are
inconsistent with the soft X-ray spectral properties of most of our
sources. ADAF models (e.g. Quataert \& Narayan 1999) further predict
that the X-ray luminosity is about 1-2\,dex higher than the radio
luminosity. For our sources however, the X-ray--to--radio luminosity
ratio is estimated to be $>10^{3}$, disfavoring the  ADAF scenario. 

Comastri et al. (2002) argue on the basis of the multiwavelength 
observations that the XBONG \chandra source 031238.9-765134 could be
associated with a completely hidden  AGN. This source presents a
flat spectrum with $\Gamma\approx 1.10\pm0.35$. The absence of a
narrow-line  emission optical spectrum typical of obscured AGN could
be explained if the source were  {\it spherically} covered by the
obscuration screen,  instead of the toroidal obscuration model which
is  usually assumed by the AGN unification models. Alternatively,
Moran et al. (2002) propose that  at higher redshifts ($z>0.1$) the
low equivalent width   narrow emission lines may be readily masked by
the  optical galaxy continuum.  The possibility that a population of
hidden AGN  is hosted in apparently normal galaxies is intriguing.
This could partially explain the scarcity of obscured AGN in
the deep \chandra surveys.  However, in our sample, we find no
significant evidence for flat X-ray spectra or equivalently large
absorbing columns. We note again however,  that our sample may be
biased against such  heavily obscured sources, as it is selected in
the total 0.5-8 keV band. Indeed, as the effective area of \xmm is
high at soft energies,  the total energy band contains a large  number
of soft sources and hence, the {\it fraction} of absorbed XBONG may be
low. Our sample is further biased  against high $f_X/f_{opt}$ objects
as it is selected to  contain optically bright sources with
spectroscopic information available in SDSS. We nevertheless believe
that the latter does not introduce any bias  against obscured XBONG. At
relatively low redshifts ($z\approx0.4$) a large  absorbing column
will reduce the X-ray emission relative to the optical and results
in low $f_X/f_{opt}$ values; for example the source 031238.9-765134
(Comastri et al. 2002) has $\log f_X / f_{opt} \approx -1$. 
Interestingly, the hard to soft band flux ratios  of many XBONG in the
sample of Comastri et al. (2003), especially those selected in the
\xmm fields, are consistent with soft X-ray spectra (see their
Fig. 4).  

\section{conclusions}
 We explore the X-ray properties of a sample of 12 X-ray bright
 optically inactive galaxies (XBONG). These are detected in 20 \xmm
 fields in the total 0.5-8 keV band overlapping with the SDSS. We
 concentrate only on those systems with available SDSS optical
 spectroscopic information and select sources which present only
 absorption lines in their optical spectra. We further select our
 objects to have high X-ray to optical flux ratios 
 ($f_{X}/f_{opt} > -2$) to reduce contamination by normal 
 galaxies. The resulting sample covers the luminosity range  $L_X
 (0.5-8 \rm \, keV)= 5 \times 10^{41}  - 2 \times 10^{44}\, erg  \,
 s^{-1}$ and  the redshift interval $0.06 < z < 0.45$. Our sample
 comprises  (i) extended X-ray sources  most probably associated with 
 galaxy clusters,  (ii) normal galaxies and (iii) {\it unobscured}
 AGN.  The unobscured AGN do not present emission lines probably
 because the optical light from the nucleus is diluted by a strong
 galaxy component (e.g. Moran et al. 2002; Severgnini et al. 2003, 
 Georgantopoulos et al. 2003). Previous work (Comastri et al. 2002) has suggested that
 the lack of optical emission  lines in XBONGs could be attributed to
 the fact that the central source suffers from large obscuration. We
 find no evidence in our sample for the presence of XBONGs which
 present significant X-ray absorption. As we imposed a cut-off in
 optical magnitude, our sample is not complete and therefore we cannot
 conclusively rule out the possibility that some  XBONGs fainter  
 that our magnitude limit present X-ray absorption. Nevertheless, our
 present work shows that the absence of optical emission lines in at
 least a  fraction of XBONGs, can be explained from the dilution  
 rather than absorption of the optical nuclear light.

\section{acknowledgments}
This work is jointly funded by the  European Union and the Greek
Ministry of Development in the framework of the Programme
'Competitiveness -- Promotion of Excellence in Technological
Development and Research-- Action 3.3.1', Project 'X-ray Astrophysics
with ESA's mission XMM', MIS-64564. 

Funding for the creation and distribution of the SDSS Archive has
been provided by the Alfred P. Sloan Foundation, the Participating
Institutions, the National Aeronautics and Space Administration, the
National Science Foundation, the U.S. Department of Energy, the
Japanese Monbukagakusho, and the Max Planck Society. The SDSS Web
site is http://www.sdss.org/. The SDSS is managed by the
Astrophysical Research Consortium (ARC) for the Participating
Institutions. The Participating Institutions are The University of
Chicago, Fermilab, the Institute for Advanced Study, the Japan
Participation Group, The Johns Hopkins University, Los Alamos
National Laboratory, the Max-Planck-Institute for Astronomy (MPIA),
the Max-Planck-Institute for Astrophysics (MPA), New Mexico State
University, University of Pittsburgh, Princeton University, the
United States Naval Observatory, and the University of Washington.

\end{document}